\documentclass[journal]{IEEEtran}
\usepackage{graphicx}
\begin{document}
\title{Transparent Microwave Photonic Filter Based on Brillouin Losses in Optical Fiber}
\author{Cheng~Feng,
        Stefan~Preu\ss ler,
        and~Thomas~Schneider
\thanks{C. Feng, S. Preu\ss ler and T. Schneider are with the Institut f\"ur Hochfrequenztechnik,
Technische Universit\"at Braunschweig, 38106 Braunschweig, Germany. (e-mail: cheng.feng@ihf.tu-bs.de;
stefan.preussler@ihf.tu-bs.de; thomas.schneider@ihf.tu-bs.de)}
\thanks{Manuscript received XXXX XX, 20XX; revised XXXX XX, 20XX.}}

\markboth{JOURNAL OF LIGHTWAVE TECHNOLOGY,~Vol.~X, No.~X, XXXX~20XX}%
{Shell \MakeLowercase{\textit{et al.}}: Bare Demo of IEEEtran.cls for IEEE Journals}

\maketitle
\begin{abstract}
In this paper, we propose a low-noise, bandwidth tunable microwave photonics filter (MPF) based on stimulated Brillouin scattering (SBS) losses. By suppressing the out-of-band signal with two broadened symmetric SBS losses, pass bandwidth can be tuned from 500 MHz to 10 GHz. Considering the limited interaction in the center frequency range, a flat-top response with 0.3 dB ripple is easily achieved. Unlike a SBS gain based filter, due to the transparency of the pass band in our proposed filter, hardly no noise is detected in a noise measurement against an obvious maximum 5 dB noise pedestal for a SBS gain based one. Considering the wide independent bandwidth and center frequency tunability, flat-top response, and low-noise characteristic, our proposed filter can be perfectly used as a supplement of most commericalized tunable single bandpass filters, whose minimum bandwidth is limited by 10 GHz. 
\end{abstract}


\IEEEpeerreviewmaketitle

\section{Introduction} 
\IEEEPARstart{D}{uring} the past decades, there was an increasing interest in realizing a flexible signal filtering function with photonic devices, which in general are named as microwave photonics filter (MPF). Especially in dense wavelength-devision-multiplexed networks, MPF is playing a more and more important role due to its excellent performance of low loss, immunity to electromagnetic interference, tunability and reconfiguratability \cite{Capmany2006}. Despite the already commericalized fiber Bragg gratings (FBG) technique, to limit the bandwidth below 10 GHz, while maintain the center frequency tunability is still challenging. On the other hand, although the well behaved filters based on Mach-Zehnder interferometer (MZI) and Fabry-Perot interferometer (FPI) can easily reach a narrow bandwidth in MHz range, their periodic pass bands bring also limitations \cite{Sadot1998}. All these factors are pushing the researchers to find an ideal narrow-banded aperiodic band pass MPF.

Stimulated Brillouin scattering (SBS) in an optical fiber can be classically described as an interaction between a
pump wave and a counter-propagating frequency downshifted Stokes wave via an acoustic wave \cite{Kobyakov2010a}. It owns so much advantages, such as low threshold, high gain, ultra narrow natural bandwidth of $\sim$30 MHz and even down to $\sim$3 MHz with specific technique \cite{Preussler2011e,Preussler2016,Wiatrek2012c} that is ideal for the application of tunable band pass filter (BPF). Therefore, SBS gain based filter has been intensively studied in the past few years. These investigation ranges from the enhancement of an independent tunability of center frequency and bandwidth through proper modulation of the pump laser \cite{Zhu2006a,Zadok2007,Tanemura2002}, via pursuing a higher selectivity using polarization pulling \cite{Stern2014a} and multi-stage configuration \cite{Wei2015}, and a rectangular flat-top response by modulation feedback compensation \cite{Yi2016} to a polarization indepdent SBS filter. Recently, a polarization indepdent SBS gain based filter with a high selectivity, low pass band ripple and independent tunability of center frequency and bandwidth is demonstrated \cite{Yi2016}, which meets the high criteria of an ideal MPF almost in every aspect. However, as a disadvantage of almost every amplification mechanism, SBS gain also introduces amplified spontaneous emission (ASE) noise \cite{Ferreira1994}. According to a recent report, the noise level is relative high under a high pump power, which is unfortunately uncompromised with a high filter selectivity \cite{Wei2015}. Furthermore, according to the Kramer-Kronig relation, SBS gain always accompanies a corresponding phase change \cite{Zadok2006} and thus, there exists also limitations for phase sensitive applications.

In this paper, we propose a novel method to demonstrate a MPF based on SBS losses instead of gain. By suppressing the out-of-band signal with two broadened symmetric SBS losses, pass bandwidth can be tuned from 500 MHz to 10 GHz by properly broadening the pump wave and choosing the proper frequency seperation between the pump waves. Due to the transparency of pass band, a flat-top response is easy to achieve. A noise measurement shows that our proposed filter can provide almost no noise filtering while a maximum 5 dB noise pedestal is detected under the same pump power for a SBS gain based one. Considering the wide tunability, flat-top response and low-noise characteristic, our proposed filter is able to work as an almost ideal MPF in cooperation with most of the commericalized filters, whose mimimum bandwidth is limited by 10 GHz. 

\section{Principle}
The aim of our proposal is to free the SBS gain filter from the ASE noise while remains the other key parameters still in high quality within 10 GHz range. Therefore, our proposed filter is working in an SBS loss mechanism instead of gain. As shown in Fig. 1, in order to circumvent the probelm of four-wave-mixing due to the multi-tones \cite{Wei2015a}, two pump waves (pump A \& B) are modulated by a same sweeping signal, whose sweeping periodicity is within the signal propagation time in the fiber \cite{Yi2016}. In the Brillouin frequency upshifted range, both sides of the input broad signal are suppressed symmetrically by the broadened SBS loss (loss A \& B) due to the pump modulation while the center frequency part transmits through the pass band with transparency. 
\begin{figure}[ht!]
\centering
\includegraphics[width=8cm]{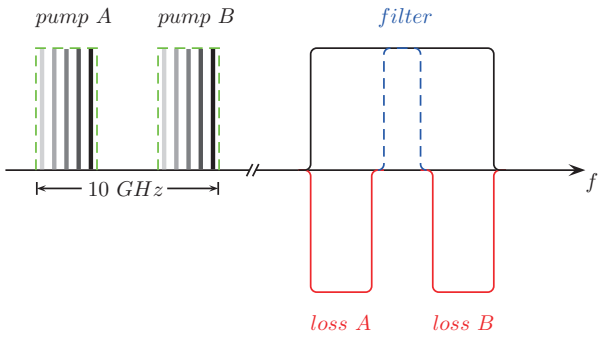}
\caption{Principle of proposed filter based on symmetric SBS losses}
\end{figure}
By carefully modulating the frequency seperation between two pumps and controlling the broadened bandwidth of pump waves, the total pump range and thereby the loss range can be well under control at approximately 10 GHz. Tuning the pump frequency and the amplitude of sweeping modulation signal, filter center frequency and bandwidth can be tunned arbitrarily und independently. Due to the transparency in the pass band, flat-top response and low-noise characteristic are easy to achieve. Thus, our proposed filter is in principle able to work as an ideal MPF within a specfic frequency range. 

\section{Bandwidth tunability}
The experimental setup of our proposed filter is illustrated in Fig. 2. A distributed feedback (DFB) laser diode (LD 1) operating at $\sim$1549.26 nm is broadened by current direct modulation by a 50 kHz ramp signal from a signal generator (SG). The broadened linewidth depends on the modulation amplitude. With a proper driving bias voltage, a radio function generator (RFG 1) modulates the CW light from LD 1 to generate a tripled signal with the power of sidebands equaling to the carrier \cite{Soto2013d}. The modulation frequency of RFG 1 and SG are selected so that the broadened carrier, high frequency side band and the frequency seperation between them cover a frequency range of approximately 10 GHz. An Erbium-doped fiber amplifier (EDFA 1) in the pump branch amplifies this tripled signal to 17 dBm at constant power mode and 
\begin{figure}[h!]
\centering
\includegraphics[width=8.5cm]{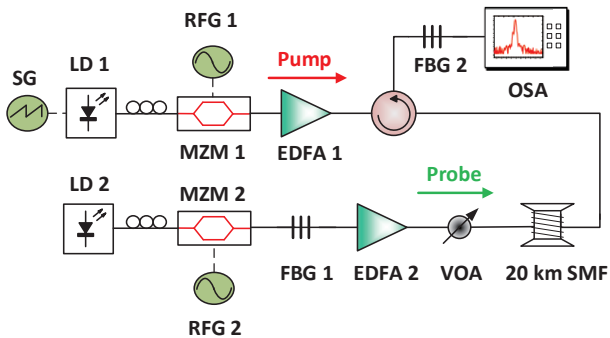}
\caption{Experimental setup. SG: signal generator, LD: laser diode, RFG: radio frequency generator, MZM: Mach-Zehnder modulator, EDFA: Erbium-doped fiber amplifier, FBG: fiber Bragg grating, VOA: variable optical attenuator, SMF: single mode fiber, OSA: optical spectrum analyzer.}
\end{figure}
emits via a circulator into the 20 km long single mode fiber (SMF) as the pump wave. On the other side, another LD (LD 2), which operates $\sim$20GHz upshifted to LD 1, is modulated by RFG 2 and works in carrier-suppressed mode. FBG 1 filters the low frequency sideband out as the probe signal. Considering that a relative high probe power works better in SBS loss regime, EDFA 2 and a variable optical attenuator (VOA) are utilized to ensure the probe power to be 5 dBm. Since the Brillouin frequency shift (BFS) of the 20 km SMF in the setup is 10.861 GHz, a scanning range of RFG 2 from 6 GHz to 20 GHz every 25 MHz is enough to get the filter profile from the peak value in optical spectrum analyzer (OSA) after the possible Rayleigh backscattering of pump wave and other frequency compoment being filtered out by FBG 2. 

In order to show the tunability of the proposed filter, experiments with a variable pass bandwidth are carried out. Fig.3 illustrates the pump signal measured by a heterodyne detection with an local oscillator (LO, not mentioned in Fig.2) in a 50 GHz photodiode (PD). By fixing the direct modulation frequency and increasing the amplitude of pump wave from SG, pump bandwidth from LD 1 increases from 0.97 GHz via 1.51 GHz and 3.41 GHz to 5.4 GHz. Despite the unflatness due to the parasitics \cite{Tucker1985}, the amplitude of the pump waves are of the same magnitude. 
\begin{figure}[h!]
\centering
\includegraphics[width=7.5cm]{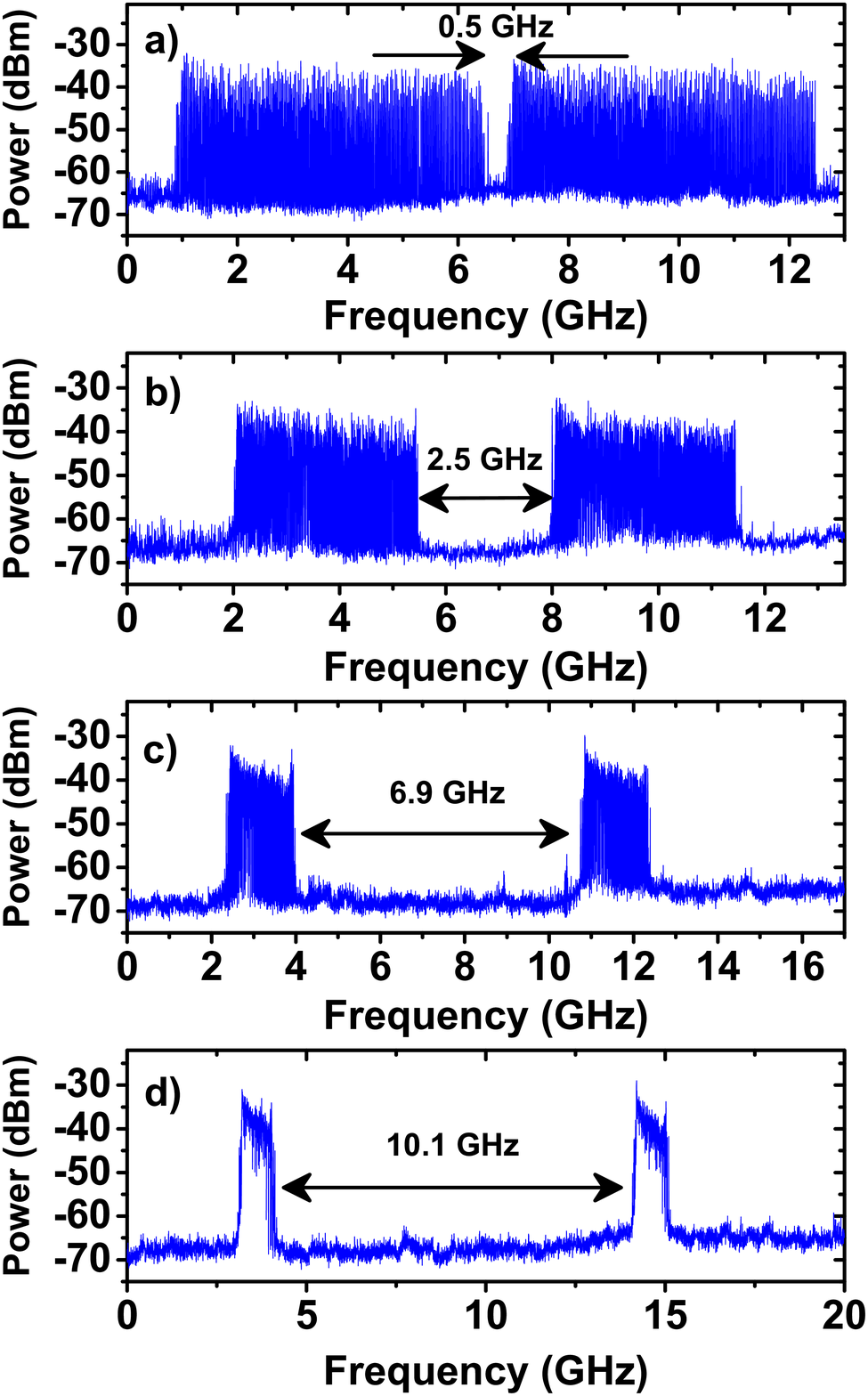}
\caption{Pump signal for the filter pass bandwidth of a) 500 MHz, b) 2.5 GHz, c) 6.9 GHz and d) 10.1 GHz}
\end{figure}
\begin{figure}[ht!]
\centering
\includegraphics[width=8cm]{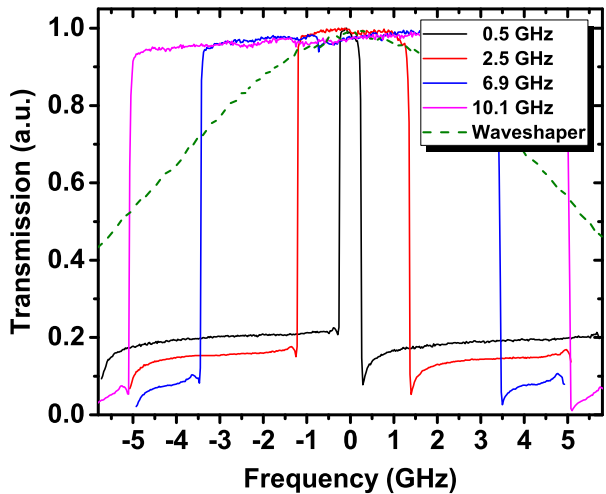}
\caption{Normalized transmission for filter pass bandwidth of 500 MHz, 2.5 GHz, 6.9 GHz and 10.1 GHz (solid lines) in comparison with commericalized WaveShaper 1000s programmable optical filter from Finisar for 10 GHz pass band (dash line).}
\end{figure}
As the normalized transmission spectrum shown in Fig.4, the filter pass bandwidth from 10.1 GHz via 6.9 GHz and 2.5 GHz to 500 MHz with ripples as low as 0.3 dB are achieved correspondingly. Note that, although the signal amplitude in the pass band is almost indepedent of the filter bandwidth, the selectivity is bandwidth dependent. According to Fig.4, generally speaking, the narrower the pass bandwidth is, i.e. the broader the pump linewidth is, the lower the selectivity is (13 dB for 10.1 GHz bandwidth while 7 dB for 500 MHz). 

Furthermore, a comparison of the transmission spectrum with a commericalized filter (Finisar WaveShaper 1000s) set at the minimum pass bandwidth (10 GHz) is made. As in Fig.4 obviously shown, due to the ultra-narrow SBS natural bandwidth, the proposed SBS loss filter provides much sharper filter edges and thus much rectangular-like filter response. 

\section{Noise measurement}
Another characteristic of the proposed filter is the transparency and low noise in the pass band. Heterodyne signals between LO and probe signal inside the filter pass band are able to compare the noise level between SBS gain based and loss based filter \cite{Wei2015}. In principle, the same setup as Fig.2 with the frequency of LD 2 downshifted for $\approx$ 20 GHz can be directly used as a SBS gain based filter. In order to make a fair comparison, the pass bandwidth for both cases are set to be the same (2 GHz) and the selectivity are selected also to be the same, i.e. SBS gain equlas SBS loss ($\approx$ 9 dB as measured in Fig.4). Considering the different mechanism of Brillouin gain and loss, different probe power is applied (-16 dBm for gain filter, while for loss case remains 5 dBm) in order to satisfy this condition under the same 17 dBm pump power. As Fig.5a shows, probe signal within the pass band of a loss based filter are well overlapped for the case when pump is on and off, indicating a full transparency of the filter pass band. However, as referred in the inset of i) of Fig.5b, despite the same SBS gain, evident pump-spectrum-like (shown in inset ii) of Fig.5b) 
\begin{figure}[h!]
\centering
\includegraphics[width=8cm]{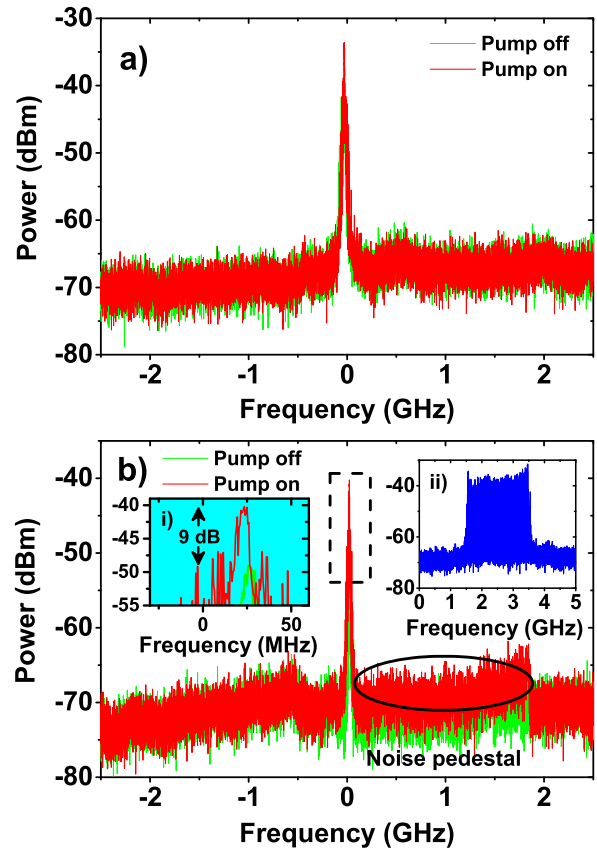}
\caption{Heterodyne signal of a local oscillator with probe signal within the pass band of a) SBS loss based filter and b) SBS gain based filter. }
\end{figure}
noise pedestal of maximum 5 dB is detected when the pump is on while vanishes when turning the pump off, which shows aggrement with the previous noise measurement of a gain based filter \cite{Wei2015}. The Brillouin noise pedestal is off-center to the right in Fig.5b due to the frequency detuning between pump and signal waves aligning at the edge in order to achieve the same selectivity. 

\section{Discussion and conclusion}
In spite of the successful demonstration of the proposed SBS loss based filter, imperfect key parameters such as limited selectivity and flat-top response remain to be improved. Fortunately, numerious priliminary works have offered several solutions like multi-stage configuration \cite{Wei2015} and polarization pulling for enhacing the selectivity \cite{Stern2014a}, and external modulation \cite{Yi2016,Wei2015a} for getting rid of parastic response due to direct modulation and get a flat-top response and a rectangular filter profile. As long as a more rectangular like filter profile is achieved, pass bandwidth can be further narrowed below 500 MHz. Another further improvement can be done by implementing the optical single side band (OSSB) modulation \cite{Yi2016} and erase the out-of-interest sideband. This could better help the proposed filter to cooperate with the most of commericalized filters. However, all these above mentioned improving methods require further investigation. 

In conclusion, we have proposd a novel bandwidth tunable MPF based on SBS losses. With the out-of-band signals being suppressed by two symmetric broadened SBS losses, pass bandwidth can be well controlled arbitrarily. A tunability ranging from 500 MHz to 10.1 GHz is demonstrated. Since the pass band is left transparent, our proposed filter has successfully overcomed the noise performance, the uncompromised disadvantage of the already well performed SBS gain based filter. As a proof of concept, an obvious noise pedestal of maximum 5 dB appears in a heterodyne noise detection for the SBS gain filter while almost no noise pedestal shows up for the loss case. Due to the flat-top response, low-noise performance, independent bandwidth and center frequency tunability within 10 GHz range, the proposed filter is able to cooperate with most of the commericalized filters and works as a ideal filter in a wider frequency range.

\section*{Acknowledgment}
The authors wish to acknowledge the financial support of the German Research Foundation (DFG SCHN 716/13-1).

\ifCLASSOPTIONcaptionsoff
  \newpage
\fi




%





\end{document}